\documentclass[aps,pra,showkeys,twocolumn,superscriptaddress]{revtex4-2}
\usepackage{array}
\usepackage{booktabs}
\usepackage{tabu}
\usepackage{dcolumn}
\usepackage{floatrow}
\usepackage{amsmath}
\usepackage{amsfonts}
\usepackage{float}
\usepackage{amssymb}
\usepackage{graphicx,color}
\usepackage[colorlinks={true}]{hyperref}
\hypersetup{colorlinks=true,linkcolor=blue,citecolor=blue,urlcolor=blue}
\usepackage{graphicx}% Include figure files
\usepackage{dcolumn}% Align table columns on decimal point
\usepackage{bm}% bold math
\usepackage{pstricks}
\usepackage{braket}
\usepackage{orcidlink}
\usepackage{lipsum}
\def\be{\begin{equation}}
  \def\ee{\end{equation}}
\def\bea{\begin{eqnarray}}
\def\eea{\end{eqnarray}}
\def\f{\frac}
\def\n{\nonumber}
\def\l{\label}
\def\p{\phi}
\def\o{\over}
\def\R{\rho}
\def\pa{\partial}
\def\om{\omega}
\def\na{\nabla}
\def\P{\Phi}
\begin{document}

\title{Remote Charging of Quantum Batteries via a Multimode Resonator} 

\author{Maryam Hadipour \orcidlink{0000-0002-6573-9960}}
\affiliation{Faculty of Physics, Urmia University of Technology, Urmia, Iran}

\author{Soroush Haseli \orcidlink{0000-0003-1031-4815}}\email{soroush.haseli@gmail.com}
\affiliation{Faculty of Physics, Urmia University of Technology, Urmia, Iran}
%\affiliation{School of Physics, Institute for Research in Fundamental Sciences (IPM),  P.O. Box 19395-5531, Tehran, Iran}

\date{\today}% It is always \today, today,
\def\be{\begin{equation}}
  \def\ee{\end{equation}}
\def\bea{\begin{eqnarray}}
\def\eea{\end{eqnarray}}
\def\f{\frac}
\def\n{\nonumber}
\def\l{\label}
\def\p{\phi}
\def\o{\over}
\def\R{\rho}
\def\pa{\partial}
\def\om{\omega}
\def\na{\nabla}
\def\P{$\Phi$}

%%%%%%%%%%%%%%%abstract%%%%%%%%%%%%%%%
\begin{abstract}
We study the remote charging of a qubit battery by a qubit charger coupled through a finite multimode resonator. The charging protocol uses time-shifted coupling pulses that account for the spatial phases of the resonator modes. We compare the energy stored in the battery with its ergotropy, the maximum work extractable by a local unitary operation. In the single-excitation regime considered here, the battery stores energy at any nonzero excited-state population, but its ergotropy is positive only when that population exceeds one half. In the absence of loss, stronger coupling brings the battery above this threshold earlier, while the final charge remains nearly complete across the coupling range studied. For the parameters considered, qubit relaxation suppresses ergotropy more strongly than resonator loss at comparable decay rates. After the battery is disconnected, relaxation causes its stored energy to decay exponentially, yet its ergotropy vanishes at a finite time while half of its energy capacity remains occupied. For a fixed storage interval, disconnecting at the population maximum therefore preserves the most extractable work.
\end{abstract}


\maketitle

%%%%%%%%%%%%%%%%%%%%%%%%%%%%%%%%%%%%%%%%%%
%%%%%%%%%%%%%%%%%%%%%%%%%%%%%%%%%%%%%%%%%%
\section{Introduction}
Quantum batteries provide a setting for studying how energy stored in a quantum system can subsequently be extracted as work~\cite{ref1, ref2,Ahmadi,Tabesh2020,Kamin2024,Zakavati2021Bounds,Hadipour2024Cavity,Hadipourneq}. Research in this area has examined the role of collective effects in energy absorption and work extraction~\cite{ref3, ref4}, the influence of many-body interactions on charging power~\cite{ref5, ref6}, and the limits that passivity places on extractable work~\cite{ref7, ref8}.

For a remotely charged battery, these questions are closely connected, transferring an excitation to the battery increases its energy, but does not necessarily make work available, either at the end of charging or after a period of storage. When the charger and battery are spatially separated, the channel connecting them is part of the charging dynamics. In circuit quantum electrodynamics, for example~\cite{ref9, ref10}, distant qubits can exchange excitations through a transmission line or a resonator. The fraction of the initial excitation that reaches the battery then depends on how the qubits couple to the channel and how much excitation is lost during propagation~\cite{ref11, ref12}. One challenge is that a reciprocal coupling permits excitation exchange in both directions. Population transferred to the battery can therefore return to the channel while the coupling remains active~\cite{ref13}. One way to impose a preferred direction is to use a nonreciprocal device~\cite{ref14, ref15}. Charging a battery prepared in its ground state, however, does not require the channel itself to be nonreciprocal. The relevant task is to coordinate emission by the charger with absorption by the battery and to determine when the battery should be disconnected.

A second challenge is to transfer the excitation before it is lost. Channel damping and thermal fluctuations can both affect transfer~\cite{ref16, ref17}; here we isolate the effects of channel loss and qubit relaxation by considering a channel initially in vacuum and reservoirs at zero temperature. When a single resonant mode is well separated from neighboring modes, adiabatic dark-state protocols can suppress its occupation during transfer~\cite{ref18, ref19}. The longer duration required for adiabatic evolution, however, can increase exposure to qubit relaxation. In the opposite limit of a dense mode spectrum, a continuum input--output description is appropriate when propagation delays and reflected fields can be neglected on the relevant dynamical timescales~\cite{ref20, ref21}.

Finite channels connecting remote superconducting qubits motivate the intermediate regime in which the coupling is comparable to the free spectral range~\cite{ref22, ref23}. Several modes can then participate in transfer, but their discrete frequencies and relative phases still matter. In particular, standing-wave modes of opposite parity couple with opposite relative signs at the two ends of the resonator. A single-mode model misses this interference, whereas a memoryless continuum model omits the finite propagation delay~\cite{ref24}.

In this work, we study a two-level battery charged by a distant qubit through a finite multimode resonator, using the asynchronous dark-state framework developed for quantum state transfer~\cite{ref24}. The charger and battery couplings are coordinated with the propagation delay and the spatial phases of the resonator modes. Our focus is on what this transfer accomplishes as a charging protocol: when the battery first provides extractable work, how much work is available at disconnection, and how much remains after storage. For the prescribed pulses, the resonant lossless calculations give nearly complete charging across the coupling range studied. Stronger coupling primarily advances the time at which work becomes available rather than increasing the final stored energy. Detuning the battery while keeping the resonant controls fixed reduces the transferred population and produces a larger relative reduction in extractable work than in stored energy.

To distinguish these quantities, we use the battery ergotropy, defined as the maximum work extractable by a cyclic unitary operation on the battery alone. For a battery state $\rho_B$ and Hamiltonian $H_B$, it is
\begin{equation}
\mathcal{W}_B
=
\operatorname{Tr}(\rho_B H_B)
-
\min_{U_B}
\operatorname{Tr}\!\left(
U_B\rho_B U_B^\dagger H_B
\right),
\label{eq:ergotropy_definition}
\end{equation}
where the minimization is over all battery unitaries $U_B$. The minimum is attained by assigning the eigenvalues of $\rho_B$ in decreasing order to the energy levels of $H_B$ in increasing order. The resulting state is passive: no further work can be extracted from it by a cyclic unitary operation on the battery alone~\cite{ref7}.
\section{Model and methods}\label{model}
\begin{figure}[t]
    \centering
    \includegraphics[width=\textwidth]
    {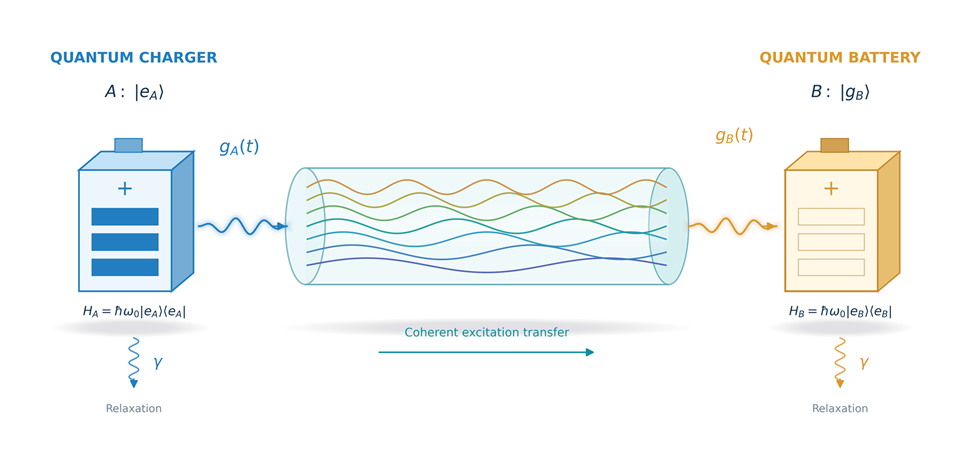}
    \caption{Schematic of the quantum battery setup. An initially
excited two-level charger $A$ supplies energy to a two-level
battery $B$, initially in its ground state, through a multimode
resonator. The time-dependent couplings $g_A(t)$ and $g_B(t)$
control the charging process. Both qubits undergo relaxation
at rate $\gamma$.
}
\label{fig1}

\end{figure}

We consider a physical setup comprising a two-level charger $A$ and a two-level battery $B$, coupled to the opposite boundaries of a finite multimode resonator. Both subsystems share an identical transition frequency $\omega_0$ and interact exclusively through the mediation of the resonator modes. Their respective raising and lowering operators are defined as $\sigma_j^+ = |e_j\rangle\langle g_j|$ and $\sigma_j^- = (\sigma_j^+)^\dagger$ for $j \in \{A, B\}$. The transmission channel is modeled by bosonic modes $a_k$ with frequencies $\omega_k$, satisfying the standard commutation relations $[a_k, a_\ell^\dagger] = \delta_{k\ell}$. The architecture and the delayed-interference
construction used below follow the multimode
state-transfer framework of Ref.~\cite{ref24}. As shown in Fig.~\ref{fig1}, an initially excited two-level
charger $A$ charges a two-level battery $B$ through a finite
multimode resonator, with the two qubits coupled to its
opposite ends.

The total Hamiltonian of the physical model is given by $ H(t) = H_0 + V(t)$, where $H_0=H_A + H_B + H_{\mathrm{ch}}$, $H_j=\hbar\omega_0 \sigma_j^+ \sigma_j^-$ and $H_{\mathrm{ch}}=\sum_{k\in\mathcal K}
    \hbar\omega_k a_k^\dagger a_k$. Within the rotating-wave
approximation, the interaction is
\begin{equation}
    V(t)
    =
    \hbar\sum_{k\in\mathcal K}
    \left[
        g_A(t)\sigma_A^+a_k
        +s_k g_B(t)\sigma_B^+a_k
        +\mathrm{H.c.}
    \right].
    \label{eq:V}
\end{equation}
The time-dependent coupling strengths between the channel and subsystems $A$ and $B$ are described by the complex envelopes $g_A(t)$ and $g_B(t)$, respectively. Although the two couplings can be controlled separately, their time
dependence must be coordinated to follow the charging
protocol. For qubits placed at opposite ends of the
resonator, we use the standing-wave phase convention
$s_k=(-1)^k$, so the relative sign of the coupling alternates
between neighboring modes. The coupling magnitudes are assumed to be independent of the mode index across the retained spectral window $\mathcal{K}$. The ideal channel spectrum is $ \omega_k=k\Delta_{\mathrm{fsr}}$. 
Here $\Delta_{\mathrm{fsr}}$ is the angular-frequency spacing
between adjacent modes. The characteristic propagation time between the two ends is given by $t_0 = \pi / \Delta_{\mathrm{fsr}}$. For a dispersionless resonator of length $L$ and wave speed $v$, we have $\Delta_{\mathrm{fsr}} = \pi v / L$, so the travel time simply becomes $t_0 = L / v$. We consider coupling strengths of the same order as the
spacing between neighboring channel modes, but much smaller
than the qubit transition frequency $g_{\max} \sim \Delta_{\mathrm{fsr}} \ll \omega_0$,
where $g_{\max}$ is the largest coupling magnitude reached
during the protocol. In this regime, several channel modes
can contribute to charging, while the rotating-wave
approximation remains appropriate.
 In this regime, several channel modes contribute to the
dynamics, and their relative phases affect the energy transfer.
A single-mode description is therefore generally insufficient.
 The rotating-wave approximation also requires the combined
coupling to the relevant modes to be much smaller than the
qubit and mode frequencies. In addition, the coupling
envelopes must change slowly compared with the rapid
oscillations at these frequencies.
 Finally, we must check that including more channel modes
does not significantly change the numerical results.
The total number of excitations is given by the operator
\begin{equation}
    N
    =
    \sigma_A^+\sigma_A^-
    +\sigma_B^+\sigma_B^-
    +\sum_k a_k^\dagger a_k,
    \label{eq:N}
\end{equation}
which is conserved since it commutes with the Hamiltonian, $[H(t), N] = 0$. We assume that the total system initially contains exactly
one excitation. In particular, the charger is initially excited, while the
battery is in its ground state and all channel modes are
in the vacuum state. The initial energy stored in the charger is simply provided by the excited state, assuming its preparation relies on an external resource. The energy cost of this preparation is left out of our charging model and would need to be accounted for separately in a full cycle analysis. 
 In the absence of dissipation, the dynamics are strictly confined to the single-excitation sector. In the interaction picture with respect to the free
qubit and channel Hamiltonians, and within the
rotating-wave approximation, the interaction Hamiltonian
is, with $\hbar=1$,
\begin{align}
H_I(t)
={}&g_A(t)\sigma_A^{+}
\sum_k a_k e^{-i\delta_k t}
\nonumber\\
&+g_B(t)\sigma_B^{+}
\sum_k(-1)^k a_k e^{-i\delta_k t}
+\mathrm{H.c.},
\label{eq:interaction_hamiltonian}
\end{align}
where $\delta_k=\omega_k-\omega_0$.

The interaction transfers the excitation between the charger,
battery, and channel without creating or removing excitations.
Thus, with the charger initially excited, the battery in its
ground state, and the channel in the vacuum state, the system
remains in the single-excitation sector as long as there
are no losses.

\begin{align}
    |\psi_I(t)\rangle
    ={}&
    \alpha(t)|e_Ag_B,\mathbf 0\rangle
    +\beta(t)|g_Ae_B,\mathbf 0\rangle
    \nonumber\\
    &+
    \sum_k c_k(t)|g_Ag_B,1_k\rangle,
    \label{eq:psi}
\end{align}
where $|1_k\rangle = a_k^\dagger|\mathbf 0\rangle$ represents the single-photon state of mode $k$. The initial conditions at $t=t_i$ are $\alpha(t_i)=1$, $\beta(t_i)=0$ and $c_k(t_i)=0$.
The charging objective is to transfer the initial
excitation to the battery.

The aim of charging is to move the initial excitation from
the charger to the battery. Charging is complete when
$|\beta(t_{\mathrm{end}})|^2=1$, meaning that the battery
is fully excited, the charger is in its ground state,
and the channel contains no photons.
\subsection{Delayed control and the asynchronous
dark-state condition}
The equal spacing between channel modes gives a travel time
of $t_0=\pi/\Delta_{\mathrm{fsr}}$ for a wavepacket to move
from one end of the resonator to the other. A full round
trip takes $2t_0=2\pi/\Delta_{\mathrm{fsr}}$.
 Due to the time needed for the field to travel between the
qubits, their interaction is delayed. The propagation delay means that the qubits do not affect each other immediately. So, the signal carried by the field takes time to travel between the qubits. After this delay, the mode phases with opposite signs
for neighboring modes make the battery couple to the same
combination of channel modes that the charger coupled to
earlier, apart from a common phase factor. For equally spaced modes with $\omega_k=k\Delta_{\mathrm{fsr}}$,
the delay $t_0$ is the time needed for a signal to travel
between the two qubits. Taking this travel time into account, the shifted battery
amplitude can be written as $ \widetilde{\beta}(t)=\beta(t+t_0)$. This allows the charger amplitude at time $t$ to be compared
with the battery amplitude at time $t+t_0$. For equally spaced mode frequencies $\omega_k=k\Delta_{\mathrm{fsr}}$,
the battery coupling shifted by the travel time is written as
\begin{equation}
    \widetilde g_B(t)
    =e^{i\omega_0t_0}g_B(t+t_0).
    \label{eq:shifted_coupling}
\end{equation}
Here, the phase factor $e^{i\omega_0t_0}$ accounts for the
phase associated with this time shift. If all mode frequencies are shifted by the same amount,
the extra constant phase can be included in the definition
of the shifted coupling $\widetilde{g}_B$.
 We take the coupling strengths $g_A(t)$ and $\widetilde{g}_B(t)$ to be active only within a finite time window $[t_i, t_f]$, vanishing everywhere outside it. Therefore, the control applied to the battery ends at
$t_f+t_0$. Using the Schr\"odinger equation with the interaction
Hamiltonian in Eq.~(\ref{eq:interaction_hamiltonian}) and the state in Eq.~(\ref{eq:psi}), the equations
of motion for the charger, battery, and channel-mode will be obtained.  Solving the equations for the channel modes gives the final amplitude of mode $k$ at time $t_f+t_0$ as
\begin{equation}
c_k(t_f+t_0)
=
-i\int_{t_i}^{t_f}d\tau\,
\left[
g_A^*(\tau)\alpha(\tau)
+\widetilde g_B^*(\tau)\widetilde\beta(\tau)
\right]e^{i\delta_k\tau},
\label{eq:final_channel_amplitude}
\end{equation}
To ensure that all channel modes return to the vacuum state by the end of the protocol, we can rely on the asynchronous dark-state condition (ADSC) introduced in Ref.~\cite{ref24}
\begin{equation}
g_A^*(t)\alpha(t)
+\widetilde g_B^*(t)\widetilde\beta(t)=0.
\label{eq:adsc}
\end{equation}
This relation connects the charger state at time $t$ to
the battery state at the later time $t+t_0$, rather than
at the same time. It ensures that no excitation remains
in the channel at the end of the protocol. For the battery
to be fully charged, however, no excitation must remain
in the charger either. When the asynchronous dark-state condition (ADSC) is
satisfied, the channel amplitudes take the simple form
\begin{equation}
c_k(t)
=
-i\int_{t-t_0}^{t}d\tau\,
g_A^*(\tau)\alpha(\tau)e^{i\delta_k\tau},
\label{eq:channel_memory}
\end{equation}
where contributions outside the time interval when the
charger coupling is active are set to zero.
 By defining the channel kernel as
\begin{equation}
K(t-\tau)=\sum_k e^{-i\delta_k(t-\tau)},
\label{eq:kernel}
\end{equation}
the equations governing the qubit amplitudes take the form
\begin{align}
\dot{\alpha}(t)
={}&-g_A(t)\int_{t-t_0}^{t}d\tau\,
K(t-\tau)g_A^*(\tau)\alpha(\tau),
\label{eq:charger_dynamics}
\\
\dot{\widetilde\beta}(t)
={}&\widetilde g_B(t)\int_t^{t+t_0}d\tau\,
K(t-\tau)
\widetilde g_B^*(\tau)\widetilde\beta(\tau).
\label{eq:battery_dynamics}
\end{align}
Once the asynchronous dark-state condition is applied, these equations become formally decoupled, though the two control functions remain linked through Eq.~\eqref{eq:adsc}. The influence of the channel is strictly confined to time intervals of length $t_0$. In particular, Eq.~\eqref{eq:channel_memory} indicates that the channel can temporarily hold excitations during the process, even if it returns to a completely empty state by the end of the protocol. This delayed-control approach forms the foundation for transferring the excitation from the charger to the distant battery.
\subsection{ADSC-based construction of charging pulses}
\label{sec:general_solutions}
Following He and Zhang~\cite{ref24}, we first choose
the charger control and then use the amplitude equations
under the ADSC to determine the battery control. We first choose the charger control $g_A(t)$ and solve
Eq.~\eqref{eq:charger_dynamics} with $\alpha(t_i)=1$.
The control should leave very little excitation in the
charger at the end, so that $|\alpha(t_f)|^2\ll 1$. The battery control is then found using the asynchronous
dark-state condition.  Once $g_A(t)$ and $\alpha(t)$ are
known, we define
\begin{equation}
    x(t)=g_A^*(t)\alpha(t).
    \label{eq:charging_x}
\end{equation}
The asynchronous dark-state condition can then be written as
$\widetilde g_B^*(t)\widetilde\beta(t)=-x(t)$.
\begin{equation}
\widetilde g_B^*(t)\widetilde\beta(t)=-x(t).
\label{eq:charging_constraint}
\end{equation}
By defining another auxiliary function
\begin{equation}
y(t)=\int_t^{t+t_0}d\tau\,
K(t-\tau)x(\tau),
\label{eq:charging_y}
\end{equation}
where $x(t)$ vanishes outside the window $[t_i, t_f]$, the equation of motion for the battery amplitude takes the simple form $\dot{\widetilde\beta}(t)
=-\widetilde g_B(t)y(t)$. 
\begin{equation}
\dot{\widetilde\beta}(t)
=-\widetilde g_B(t)y(t).
\end{equation}
Using the above relation together with Eq.~\eqref{eq:charging_constraint}
gives
\begin{equation}
    \widetilde\beta^*(t)\dot{\widetilde\beta}(t)
    =x^*(t)y(t).
    \label{eq:charging_identity}
\end{equation}
Battery amplitude can be written as
$\widetilde{\beta}(t)=\sqrt{P_B(t)}e^{i\phi(t)}$,
where $P_B(t)$ is the excitation probability and $\phi(t)$
is the phase. Taking the real and imaginary parts of
Eq.~\eqref{eq:charging_identity} gives the equations
for how these two quantities change with time
\begin{align}
\dot P_B(t)
&=2\operatorname{Re}\!\left[x^*(t)y(t)\right],
\label{eq:charging_population}\\
P_B(t)\dot\phi(t)
&=\operatorname{Im}\!\left[x^*(t)y(t)\right].
\label{eq:charging_phase}
\end{align}
Since the battery starts out completely empty, we integrate Eq.~\eqref{eq:charging_population} over time to find the population at any time
\begin{equation}
P_B(t)=2\int_{t_i}^{t}ds\,
\operatorname{Re}\!\left[x^*(s)y(s)\right].
\label{eq:charging_population_integral}
\end{equation}
When the battery excitation probability is nonzero,
Eq.~\eqref{eq:charging_phase} gives its phase apart from
an arbitrary constant. Using these results, the battery control can be written as
\begin{equation}
\widetilde g_B(t)
=-\frac{x^*(t)}{\widetilde\beta^*(t)}
=-\frac{x^*(t)e^{i\phi(t)}}{\sqrt{P_B(t)}}.
\label{eq:charging_receiver}
\end{equation}
This expression applies when $P_B(t)>0$.
Using the time-shift convention introduced above,
the physical control reads
\begin{equation}
g_B(t)
=e^{-i\omega_0t_0}\widetilde g_B(t-t_0).
\label{eq:charging_physical_control}
\end{equation}
This method relies on the resulting control pulse being physically realizable in practice. In particular, the excitation probability $P(t)$ must stay
between zero and one. When $P(t)$ approaches zero, the ratio
in Eq.~\eqref{eq:charging_receiver} must have a finite limit. So, not every charger pulse gives a battery control
that stays finite and can be used in an experiment. Finally, we check the control pulses by solving the full
Schr\"odinger equation for the charger, channel, and battery.
If the ADSC is satisfied exactly, no excitation remains in
the channel at the final time $t_f+t_0$. In the absence of
losses, the total excitation is conserved, so the final
battery excitation probability is
\begin{equation}
|\beta(t_f+t_0)|^2=1-|\alpha(t_f)|^2.
\label{eq:charging_completion}
\end{equation}
Thus, if no excitation remains in the charger, the battery
is fully charged. For the numerical simulations, we set \(t_i=0\) and prescribe the
charger-channel coupling as
\begin{equation}
    g_A(t) =
    \begin{cases}
        g\bigl[1-\cos(\pi t/t_f)\bigr], & 0 \leq t \leq t_f, \\
        0,                              & \text{otherwise},
    \end{cases}
    \label{eq:charger-pulse}
\end{equation}
where \(t_f=10(2\pi/\Delta_{\mathrm{fsr}})\) is the pulse duration.
The parameter \(g\) sets the coupling scale, not the maximum of coupling. 
The pulse reaches \(g\) at \(t=t_f/2\) and \(2g\) at \(t=t_f\),
after which the coupling is switched off. With this prescribed
pulse, we solve Eq.~(13) for the charger amplitude subject to
\(\alpha(0)=1\).

%\begin{figure*}[t]

\begin{figure}[t]
    \centering
    \includegraphics[width=0.75\textwidth]
    {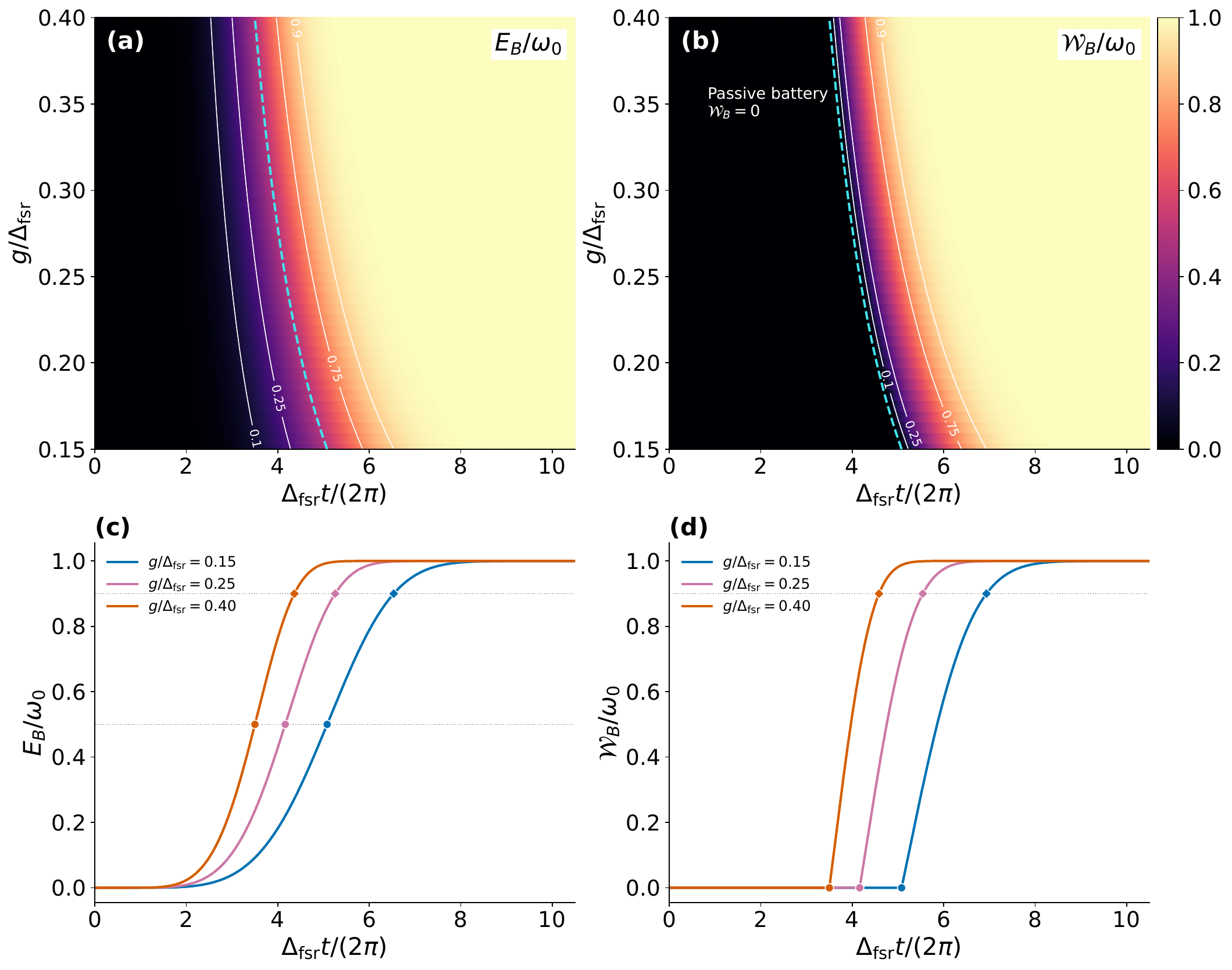}
    \caption{
Lossless charging dynamics of a battery coupled to a charger
through seven channel modes $k=-3,\ldots,3$. (a) and (b) show density
plot of the normalized stored energy
$E_B/\omega_0$ and ergotropy
$\mathcal{W}_B/\omega_0$ versus
$\Delta_{\mathrm{fsr}}t/(2\pi)$ and
$g/\Delta_{\mathrm{fsr}}$. (c) and (d) show the dynamics of $E_B/\omega_0$ and $\mathcal{W}_B/\omega_0$ respectively for different values of $g/\Delta_{\mathrm{fsr}}$.
}
\label{fig:battery_density}
\end{figure}
In Fig.~\ref{fig:battery_density}, we show how the coupling strength changes
the stored energy and ergotropy in our lossless seven-mode setup. The charger
is initially excited, the battery is in its ground state,
and all channel modes are in vacuum. The channel modes are evenly spaced around the qubit resonance frequency. The receiver control is delayed by half the inverse of the mode spacing. The setup is assumed to be free of qubit relaxation and channel loss. 

In Fig.~\ref{fig:battery_density}(a) and Fig.~\ref{fig:battery_density}(b), the normalized stored energy and ergotropy are shown as functions of time and coupling strength. Although both reach their maximum at late times, they behave quite
differently while the battery is charging.
 For the initial state and excitation-conserving
dynamics considered here, the reduced battery state remains
diagonal in its energy eigenbasis. The battery therefore
begins to store energy before it becomes active. At first the battery is passive because its internal energy is too low to
yield any work. Only when the excited-state population passes one half can
the stored energy be extracted by operations acting on the battery alone. The dashed line marks the time when the battery first reaches half of its energy capacity suitable for to extract work. It therefore represents the half-capacity contour in
Fig.~\ref{fig:battery_density}(a) and the boundary between
passive and active battery states in
Fig.~\ref{fig:battery_density}(b). The ergotropy is zero at this threshold and becomes positive as the battery charges beyond half of its energy capacity. As the coupling increases, this boundary moves to earlier times, showing that stronger coupling allows work to be extracted sooner under the chosen controls. The white lines mark constant levels of normalized stored energy or ergotropy. As the coupling increases, these lines shift to earlier times, showing that the battery reaches the same energy or ergotropy level sooner. The battery becomes nearly fully charged within the simulated time interval for all coupling strengths considered.
The main difference is therefore how quickly the battery reaches a given level of stored energy or extractable work, rather than how much energy it ultimately stores. Because the battery state is diagonal in the energy basis, both maps depend on the same excited-state population. The different patterns show how the stored energy differs from the work that can be extracted through unitary operations on the battery alone.

The stored-energy and ergotropy trajectories in
Fig.~\ref{fig:battery_density} show how the charging
dynamics depend on the coupling strength. Stronger
coupling allows the battery to store energy and develop
extractable work sooner, shortening the charging process.

The circular markers indicate the same activation time
in both figures. They mark the point where the battery
reaches half of its energy capacity in
Fig.~\ref{fig:battery_density}(c), beyond which work
becomes extractable in Fig.~\ref{fig:battery_density}(d).
The diamond markers show when the stored energy and
ergotropy first reach the same prescribed fraction of
their respective capacities. For every coupling strength
shown, the battery reaches this level of stored energy
before reaching the corresponding level of extractable
work. Thus, a nearly charged battery does not necessarily
provide an equally large fraction of its maximum
extractable work. Reaching that fraction requires the
battery to be charged closer to its full energy capacity. The delayed onset and steeper rise of the ergotropy
follow from its dependence on the battery population.
Once the battery is active, its normalized ergotropy
increases at twice the rate of its normalized stored
energy. This behavior reflects the conversion of the
battery state from passive to increasingly active and
does not imply an independent enhancement of the
excitation-transfer rate. Overall, the figure demonstrates
that stronger coupling accelerates energy accumulation
and battery activation under the chosen controls, while
highlighting that a high stored energy does not necessarily
imply an equally high fraction of extractable work.

\begin{figure}[t]
    \centering
    \includegraphics[width=0.75\textwidth]
    {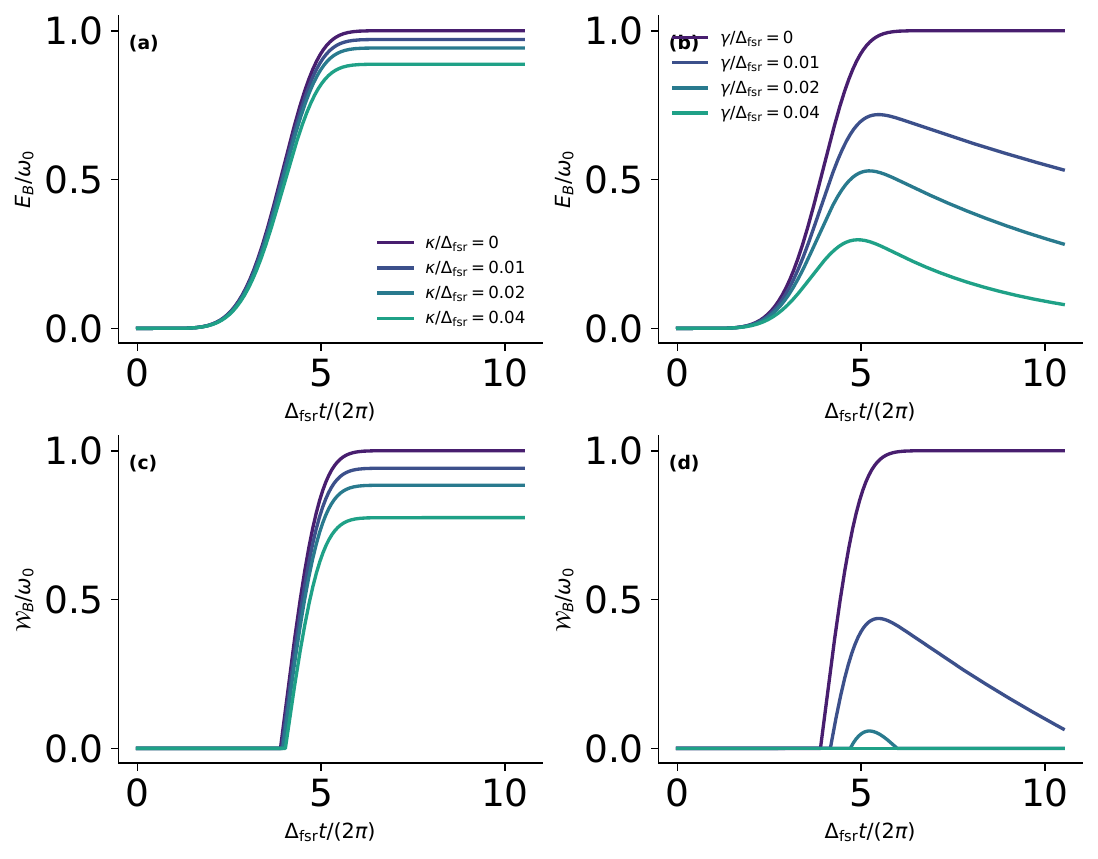}
    \caption{
Effects of dissipation on battery charging at
$g/\Delta_{\mathrm{fsr}}=0.30$ in a resonant seven-mode channel.
$E_B/\omega_0$ (top) and $\mathcal{W}_B/\omega_0$ (bottom)
are shown for different vales of channel decay $\kappa$ with $\gamma=0$
[(a),(c)] and equal different values of relaxation $\gamma$ of both qubits
with $\kappa=0$ [(b),(d)].
}
\label{fig:dissipation}
\end{figure}

\subsection{Dissipative losses, stored energy, and ergotropy}
As in the previous sections, we start with the charger in
its excited state, the battery in its ground state, and all
seven channel modes empty. The charger and the battery both relax at rate $\gamma$,
while each channel mode loses photons at rate $\kappa$.
 We assume that the losses are independent and that the
environment is at zero temperature and has no memory.
The coherent dynamics move the single excitation between
the charger, the channel, and the battery without changing
its total number. Including loss to the environment gives
\begin{equation}
P_A(t)+P_B(t)+P_C(t)+P_{\mathcal{L}}(t)=1,
\label{eq:excitation_balance}
\end{equation}
where $P_C(t)=\sum_{k=-3}^{3}\langle a_k^\dagger a_k\rangle_t$
is the probability that the excitation is in one of the seven
channel modes, and $P_{\mathcal{L}}(t)$ is the probability
that it has been lost to the environment by time $t$. The probability
that the excitation has been lost by time $t$ is
\begin{equation}
P_{\mathcal{L}}(t)
=
\int_{t_i}^{t}
\left[
\gamma\bigl(P_A(s)+P_B(s)\bigr)+\kappa P_C(s)
\right] ds.
\label{eq:loss_probability}
\end{equation}
The two terms account for excitation lost from the charger and
battery, and from the channel modes, respectively. To measure exposure to channel loss, the integrated channel
population is defined as
\begin{equation}
\mathcal{I}_C(t)
=
\int_{t_i}^{t}P_C(s)\,ds,
\label{eq:channel_exposure}
\end{equation}
To first order in the decay rates, the populations in
Eq.~\eqref{eq:loss_probability} are replaced by their lossless
values, giving
\begin{equation}
P_{\mathcal{L}}(t)
\simeq
\gamma(t-t_i)
+(\kappa-\gamma)\mathcal{I}_C(t).
\label{eq:weak_loss}
\end{equation}
For a fixed duration, a smaller integrated channel population
therefore gives less loss to leading order when $\kappa>\gamma$.
This estimate is valid only when losses are weak.
The exact energy and ergotropy relations distinguish incomplete
transfer from irreversible excitation loss. When transfer is
nearly complete, and the population remaining in the charger
and channel is negligible compared with the leading loss
correction, Eq.~\eqref{eq:weak_loss} gives
\begin{align}
\frac{E_B}{\omega_0}
&\simeq
1-\gamma(t-t_i)
-(\kappa-\gamma)\mathcal{I}_C(t),
\label{eq:energy_weak_loss}\\
\frac{\mathcal{W}_B}{\omega_0}
&\simeq
1-2\gamma(t-t_i)
-2(\kappa-\gamma)\mathcal{I}_C(t),
\label{eq:ergotropy_weak_loss}
\end{align}
where losses are assumed to be weak. These expressions show that the reduction
in normalized extractable work is twice the reduction in
normalized internal energy relative to a fully charged battery.
Thus, relaxation and channel loss reduce not only the energy
stored in the battery but also the fraction of that energy
available as work.

Fig.~\ref{fig:dissipation} shows how energy loss from
the channel and qubit relaxation affect the energy stored
in the battery and the work that can be extracted from it.
 The simulations use a resonant seven-mode channel with
the coupling parameter fixed at
$g/\Delta_{\mathrm{fsr}}=0.3$. IThe battery control
pulse follows the charger pulse with a delay of
$t_0=\pi/\Delta_{\mathrm{fsr}}$. Fig.~\ref{fig:dissipation}(a) and
Fig.~\ref{fig:dissipation}(c) show the battery energy and
ergotropy when all channel modes lose excitations at
the same zero-temperature decay rate $\kappa$, with
no qubit relaxation. Fig.~\ref{fig:dissipation}(b)
and Fig.~\ref{fig:dissipation}(d) show what happens when the charger and battery relax at the same
rate $\gamma$ and the channel is lossless $\kappa=0$.
Here, $\gamma$ applies to both qubits, not just the battery. Fig.~\ref{fig:dissipation}(a) shows that as the channel
loss rate $\kappa$ increases, the battery stores slightly
less energy at the end of charging. The shape of the
charging trajectory, however, stays nearly the same.
 The trajectories follow nearly the same path early in the charging process, then gradually level off at different
final energies set by the decay rate. At the end of the simulation, the normalized battery energy is slightly lower for stronger channel decay,
rather than reaching nearly full capacity as it does
in the lossless case. Even at the highest channel decay rate considered, the final stored energy is only slightly reduced. This weak sensitivity reflects the channel’s limited occupation during the transfer, as it briefly carries excitations from the charger to the battery rather than storing them. Channel losses can reduce the energy transferred to the
battery, but they do not directly remove energy from
the battery once it is decoupled from the channel. After the main charging phase, little excitation remains
in the channel, so further channel losses become very
small. The energy stored in the battery then stays
nearly constant. The battery energy remains nearly constant at later times
because qubit relaxation is turned off in this case,
not because the battery is protected against direct relaxation. The ergotropy trajectories in Fig.~\ref{fig:dissipation}(c)
show the same trend, with stronger channel decay leading
to less extractable work, although the relative reduction
is more pronounced than for the stored energy. Taken together, these results show that increasing the
decay rate reduces the extractable work by a somewhat
greater fraction than the stored energy. Although channel losses reduce the work that can be
extracted from the battery, a substantial amount remains
available at the end of charging in every case considered.
 By contrast, Fig.~\ref{fig:dissipation}(b) shows that
when both qubits undergo relaxation, the battery energy
initially increases, reaches a peak, and then gradually
decreases. The battery energy begins to decrease when the incoming
energy can no longer make up for the energy lost
through battery relaxation. Since both the charger and the battery relax at the same
rate, some energy is lost from the charger before it
can reach the battery, while energy already stored in
the battery also decays. Qubit relaxation therefore leaves less energy available
to charge the battery and causes the stored energy
to decrease over time. At comparable decay rates, qubit relaxation causes
a much greater loss of stored energy than channel
dissipation. At the highest rate considered, much less of the initial
energy remains stored in the battery when the loss comes
from qubit relaxation rather than channel decay. This difference occurs because energy spends only a short
time in the channel, but stays much longer in the charger
before transfer and in the battery afterward. Qubit relaxation therefore causes losses over a longer
part of the charging and storage process. As a result,
it has a much stronger effect than channel decay for
the control pulses used here, although this need not
hold for other charging protocols. Fig.~\ref{fig:dissipation}(c) and Fig.~\ref{fig:dissipation}(d) also highlight
the difference between storing energy in the battery
and being able to extract it as work through operations
on the battery alone.
 Since the battery state remains diagonal in its energy
basis, it can store energy even when no work can be
extracted from it by unitary operations on the battery alone.
 Work can therefore be extracted only when the battery
is more likely to be excited than in its ground state.
This is why the ergotropy stays zero at first, even
as the battery begins to store energy. Once the excited-state population exceeds the ground-state
population, the extractable work grows rapidly. Dissipation
reduces this population difference, so the extractable work
falls by a larger fraction than the stored energy. This effect is especially clear with qubit relaxation.
At higher relaxation rates, the battery can still store
a considerable amount of energy but provide little
extractable work, because its excited-state population
is only slightly above the threshold needed for work extraction. The ergotropy trajectories in Fig.~\ref{fig:dissipation}(d)
rise and then fall, showing that when qubit relaxation
is present, the amount of work available depends strongly
on when it is extracted. At higher relaxation rates, the extractable work reaches
a lower maximum at an earlier time. The extractable work drops quickly after reaching its maximum, so much less work is available if extraction
is delayed until the end of the pulse sequence. Although these points show the highest ergotropy along the trajectories, they do not represent an optimized extraction scheme because they do not include the feedback effects of a real extraction process. Fig.~\ref{fig:dissipation} shows how channel loss and
qubit relaxation affect the charging process separately. Channel loss mainly reduces the energy that reaches
the battery, leaving both the stored energy and the
extractable work at lower but nearly constant values.
Qubit relaxation also removes energy already stored
in the battery, so both quantities rise to a maximum value
and then decrease. This comparison shows that storing a large amount of
energy during charging does not guarantee that much
work will remain available later. For the control pulses
used here, reducing qubit relaxation helps preserve
the stored energy, while extracting work at the right
time avoids further losses.
Good charging does not necessarily imply good storage. To distinguish
the two, we consider charging the battery until $\tau_c$, disconnecting
it from the channel, and storing it for a duration $\tau_s$ before
extracting work. No work is extracted before or during storage.
After decoupling, the battery no longer exchanges energy with the
channel, but relaxation continues to reduce its excited-state
population and ergotropy.

For battery state undergoing zero-temperature relaxation,
the population decays exponentially during storage. Extractable work
remains available only while the excited-state population exceeds
the ground-state population. The time at which this condition ceases
to hold defines the active storage lifetime $t_{\mathrm{active}}$.
Beyond this time, the battery is passive. It still contains energy,
but that energy cannot be extracted as work by a unitary operation
on the battery alone.

A map of stored ergotropy versus $t_c$ and $\tau_s$ shows how the
charging stop time affects the work available after a given waiting
time. For a fixed storage duration and a constant relaxation rate,
maximizing the population at decoupling also maximizes the stored
ergotropy whenever positive work survives. A broader optimization
would require additional constraints, such as a fixed total time
for charging and storage or limits on the available controls.

\begin{figure}[t]
    \centering
    \includegraphics[width=0.75\textwidth]
    {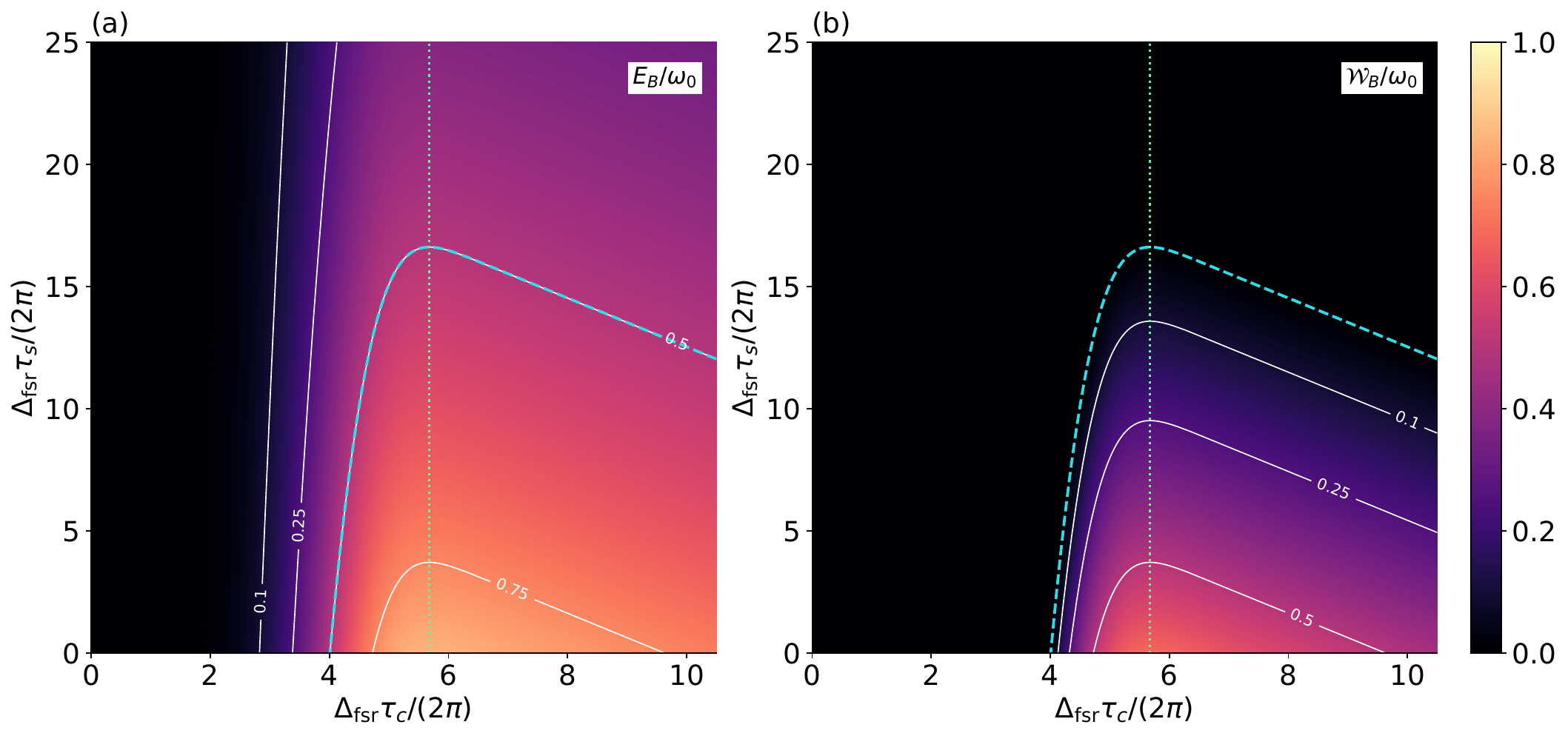}
    \caption{
Charging and storage of the seven-mode quantum battery at
$g/\Delta_{\mathrm{fsr}}=0.30$, with equal qubit relaxation rates
and no channel loss. (a) Energy and (b) ergotropy remaining after storage versus charging
stop time $t_c$ and storage duration $\tau_s$, at
$\gamma/\Delta_{\mathrm{fsr}}=0.005$ and $g/\Delta_{\mathrm{fsr}}=0.30$.
%The battery is ideally disconnected at $t_c$ and left to relax
%without work extraction. The cyan dashed boundary marks the loss
%of positive ergotropy, and the green dotted line marks the stop time
%with the highest battery population. White contours show constant
%normalized energy or ergotropy. Time is measured in units of
%$2\pi/\Delta_{\mathrm{fsr}}$, and energy and ergotropy in units of
%$\hbar\omega_0$.
}
\label{fig:charging_storage}

\end{figure}
Fig.~\ref{fig:charging_storage} show how the choice of charging stop time
$\tau_c$ affects the energy $E_B /\omega_0$ and the extractable work
$\mathcal{W}_B / \omega_0 $ that remain after storage. Each vertical line corresponds to a fixed time $\tau_c$ at which the battery
is disconnected from the channel without prior work extraction. Moving
upward along the line shows how the stored energy in Fig.~\ref{fig:charging_storage}(a) or the
ergotropy in Fig.~\ref{fig:charging_storage}(b) changes as the storage time $\tau_s$ increases. As Fig.~\ref{fig:charging_storage}(a) shows, disconnecting the battery
early leaves little energy for storage because its excitation probability
is still low.
 The green dotted line marks the time at which the battery excitation
probability $P_B$ has its maximum value. After the maximum point, the stored energy decreases even if the battery remains
connected because relaxation removes energy faster than the channel
supplies it. Once the battery is disconnected, its energy decays at the same rate
$\gamma$ for every charging stop time $\tau_c$. Disconnecting the
battery when $P_B$ has its maximum value leads to  leave the most energy after any
fixed storage time $\tau_s$.

Fig.~\ref{fig:charging_storage}(b) shows how much of the stored energy
can still be extracted as work. For the diagonal battery state considered
here, the ergotropy is positive only when the excited state is more
populated than the ground state. If the battery
is disconnected before it reaches this threshold, it can store energy
but has no extractable work, even at $\tau_s=0$. If it is disconnected at time $\tau_c$
with $P_B>1/2$, work is initially available, but the amount
decreases as the battery relaxes. The white contours connect
combinations of $\tau_c$ and $\tau_s$ that leave the same ergotropy.

The cyan dashed trajectory marks the storage time at which the excited and
ground states become equally populated. Below this trajectory, work can
still be extracted. On and above it, the ergotropy is zero even though
some energy remains in the battery. The trajectory reaches its highest point
at the green dotted line, where the excitation probability at
disconnection is largest. Stopping at this charging maximum value therefore
keeps work available for the longest time. Disconnecting earlier or
later leaves the battery closer to the threshold, so its ergotropy
vanishes sooner. The difference comes from the population at
disconnection, not from a change in the relaxation rate $\gamma$.

\begin{figure}[t]
    \centering
    \includegraphics[width=0.75\textwidth]
    {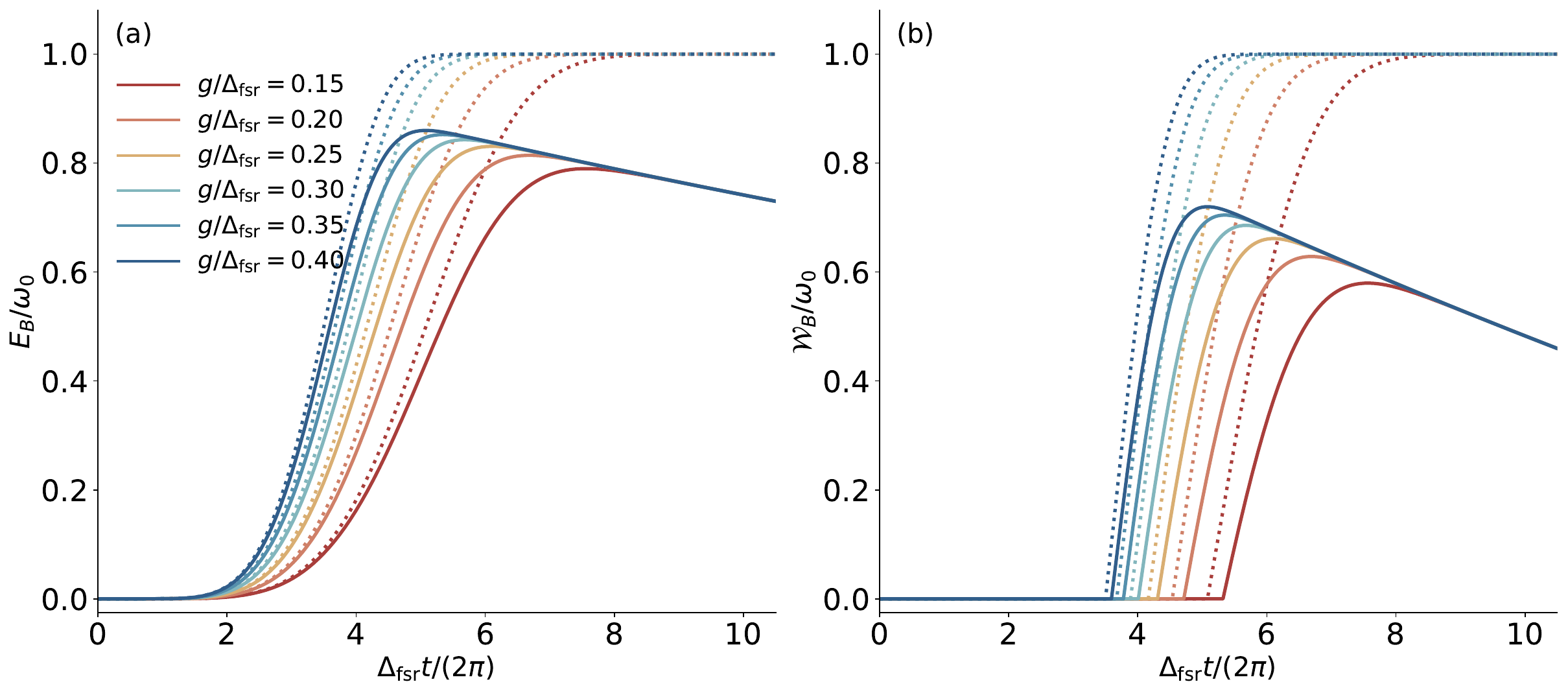}
    \caption{
(a) the normalized battery energy
$E_B/\omega_0$ and (b) the normalized ergotropy
$\mathcal{W}_B/\omega_0$ for a seven-mode channel with different values of  $g/\Delta_{\mathrm{fsr}}$. Solid trajectories include equal charger and battery decay rates $\gamma/\Delta_{\mathrm{fsr}}=0.005$. Dotted trajectories show
lossless evolution with $\gamma=0$. Channel loss is set to zero
$\kappa=0$ throughout.
}
\label{fig:coupling_relaxation}

\end{figure}
% Add \usepackage{hyperref} in your preamble
Fig.~\ref{fig:coupling_relaxation} shows how the coupling strength
and qubit relaxation affect the energy stored in the battery and
the work that can be extracted from it. 

For each coupling strength $g/\Delta_{\mathrm{fsr}}$, we use the
same control pulses with and without qubit relaxation. The solid
trajectories show the results when the charger and battery relax
at the same rate, $\gamma/\Delta_{\mathrm{fsr}}=0.005$. The dotted
trajectories show the results without relaxation, $\gamma=0$.
There is no channel loss in either case ($\kappa=0$), so any
difference between the two comes from qubit relaxation alone.

As can be seen from Fig.~\ref{fig:coupling_relaxation}(a), stronger
coupling allows the battery to charge sooner without dissipation,
while all six trajectories approach a nearly fully charged state.
 In the presence of qubit relaxation, the stored energy reaches a
peak and then decreases. Within the range studied, stronger
coupling allows the battery to reach a higher energy peak in a
shorter time.

 It can also be seen that stronger coupling leads to a higher
value of energy in a shorter time. This suggests that faster
excitation transfer gives relaxation less time to reduce the
stored energy before the battery reaches its peak.

Fig.~\ref{fig:coupling_relaxation}(b) shows that stronger coupling
also leads to maximum extractable work, although its behavior
differs from that of the stored energy at the start of charging.
  In the presence of relaxation, stronger coupling leads to a higher
maximum extractable work. The stored energy and extractable work
reach their highest values at the same time because both increase
with the battery's excited-state population whenever extractable
work is available.
At later times, the trajectories with relaxation nearly overlap.
Once excitation transfer is largely complete, the stored energy
decreases approximately exponentially as the battery continues to
relax. While extractable work remains available, its normalized
value falls faster than the normalized stored energy. Stronger
coupling therefore raises the maximum stored energy and extractable
work for the control pulses studied, but offers little benefit at
the shared final observation time. Assessing this charging protocol
thus requires considering both the highest charge reached and how
much energy remains later.
\section{Conclusion}
For a qubit battery charged through a multimode resonator, stronger coupling primarily changes when extractable work becomes available. In the absence of loss, the battery reaches nearly full charge across the coupling range studied, but stronger coupling produces positive ergotropy earlier. With qubit relaxation, stronger coupling also yields an earlier and higher ergotropy peak. By the final observation time, however, little of that advantage remains. For the pulses and decay rates considered here, qubit relaxation suppresses extractable work more strongly than loss in the resonator.

The storage dynamics show why stored energy alone is not a sufficient measure of battery performance. After disconnection, the battery energy decays gradually, but ergotropy vanishes as soon as the excited-state population falls to one half. At that point, the battery still holds half of its energy capacity, yet none of that energy can be extracted by a local unitary operation. For a fixed storage interval, disconnecting the battery when its excited-state population is highest leaves the most work available at readout. The relevant measure of charging performance is therefore not only the peak ergotropy reached during charging, but also the extractable work that remains when the battery is needed.

\end{document}